\documentclass[12pt]{article}

\usepackage{amsfonts}

\newtheorem{Proposition}{Proposition}

\begin{document}

\title{\bf Flows in nonequilibrium quantum systems\\
and quantum photosynthesis}

\author{S.V.Kozyrev, A.A.Mironov, A.E.Teretenkov, I.V.Volovich
\\
Steklov Mathematical Institute of Russian Academy of Sciences}
\date{}
\maketitle

\begin{abstract}
A three level quantum system interacting with nonequilibrium environment is investigated. The stationary state of the system is found (both for non--coherent and coherent environment) and relaxation and decoherence to the stationary state is described. The stationary state of the system will be non--equilibrium and will generate flows. We describe the dependence of the flows on  the state of the environment.

We also discuss application of this model to the problem of quantum photosynthesis, in particular, to description of flows of excitons and generation of excitonic coherences.
\end{abstract}

\section{Introduction}

In the present paper we consider a three level quantum system which interacts with three reservoirs. The reservoirs are Bose quantum fields in temperature states. Temperatures of the reservoirs are different, therefore the environment is nonequilibrium and the system interacting with the environment can be considered as an example of quantum thermodynamic machine. Moreover we will consider also the case when the state of one of the reservoirs includes also a coherent component.

We will investigate how this thermodynamic machine operates. The principal question is the form of the stationary state (which will be nonequilibrium) and the currents (or flows) in this state. The higher the flow the more effective the thermodynamics machine will be. We find the expression for the stationary state of the system and investigate the dependence of this state on the states of the reservoirs. The density matrix of the stationary state of the system for the case of non--coherent nonequilibrium environment will be diagonal (but non--Gibbs).

We introduce the generalized brightnesses of the reservoirs (proportional to the numbers of quanta of the fields which are in resonance with the corresponding transitions in the system). We show that the dependence of the flow on the brightnesses is saturated --- the flow is proportional to the brightness for small brightness and saturates (tends to constant) for large brightness.

The second question is how the obtained expressions for the stationary state of the system and the flow will be changed when one of the reservoirs contains a coherent component. We show that in this case the stationary state will be deformed, in particular it will includes off--diagonal component. The expression for the flow will also be deformed.

We also investigate convergence to the stationary state (for all discussed regimes of the environment) and describe relaxation and decoherence for the model under consideration.

For the investigation of evolution of quantum systems interacting with the environment we use the approach of the stochastic limit of quantum theory \cite{AcLuVo}, see also \cite{notes} for the discussion of quantum many particle systems in the stochastic limit. In this approach evolution of the reduced density matrix of the system will be generated by the Lindblad dissipative operator.  Stochastic limit in presence of a coherent field was discussed in \cite{lambdaatom}, in the present paper we use different but equivalent approach.

The systems which couple to multiple reservoirs are investigated in particular in  \cite{Gea}--\cite{Tru}.
For general discussion of different models in quantum thermodynamics (including heat flows) see \cite{Heat1}--\cite{Heat6}.

One of the most interesting applications of the model investigated in this paper is to the problem of quantum photosynthesis. Photosynthesis (actually, the first stages of photosynthesis) is the process of absorption of photons in light harvesting complexes with creation of excitons, transport of excitons to the reaction center and absorption of excitons in the reaction center \cite{MayKuhn}. The problem of quantum photosynthesis is related to the observed in photosyntetic systems coherences with long lifetime
\cite{Qbio}, \cite{Engel}, \cite{SFOG}. We will not discuss here this effect of coherences, see \cite{tmf2014}, \cite{darkstates} for the application to this problem of a generalization of the model described in the present paper. See also \cite{OhyaVolovich} for general discussion of quantum effects in biology.


\section{Hamiltonian and generator of evolution}\label{the_system}

\noindent{\bf The Hamiltonian of a system interacting with three reservoirs.}\quad In this article we discuss a non--equilibrium stationary state for the model of quantum system interacting with three different reservoirs with three different temperatures. In particular a model of this kind describes quantum photosynthesis. Since the model is non--equilibrium the stationary state will correspond to a flow. For the model of quantum photosynthesis this is a flow of excitons to the reaction center.

We consider a system with three non--degenerate levels and a Hamiltonian
$$
H_S=\varepsilon_0 |0\rangle\langle 0|+ \varepsilon_1 |1\rangle\langle 1|+\varepsilon_2 |2\rangle\langle 2|.
$$

In our model the system describes a photosynthetic system (in one--exciton approximation in the so called global basis), $|0\rangle$ corresponds to a state without excitons, $|1\rangle$ corresponds to a state ''exciton in the reaction center'' (or {\it sink}), $|2\rangle$ corresponds to one-exciton state of a chromophore.

We consider the following set of transitions between the energy levels. Transitions between $|0\rangle$ and $|2\rangle$ (exciton excitation and its stimulated absorption) are coupled with the interaction with light (thermal reservoir at 6000K), transitions between $|2\rangle$ and $|1\rangle$ (exciton transport to the reaction center) are coupled with the interaction with phonons, or oscillations of a protein matrix (thermal reservoir at 300K), and transitions between $|1\rangle$ and $|0\rangle$ (exciton absorption in the reaction center) are coupled with the interaction with a Fock reservoir (i.e a reservoir with zero temperature).

Thus, we have three reservoirs, described by quantum bosonic field Hamiltonians $H_{\rm em}$ (electro\-magnetic field), $H_{\rm ph}$ (phonons, oscillations of the protein matrix), $H_{\rm sink}$ (sink, or exciton absorption in the reaction center). For each reservoir the corresponding Hamiltonian takes the form
$$
H_R=\int_{\mathbb{R}^3} \omega_R(k)a^{*}_R(k) a_R(k) dk,
$$
where $R={\rm em,\, ph,\, sink}$ enumerates the fields, $\omega_R$ is a dispersion of the field $a_R$.

Each of the reservoirs $R$ is in a Gaussian mean zero state with quadratic correlation function
$$
\langle a^{*}_{R}(k)a_R(k') \rangle=N_R(k)\delta(k-k').
$$

Here $N_R(k)$ (number of field quanta with wave number $k$) equals (in a thermal state with inverse temperature $\beta_R$)
$$
N_R(k)={1\over{e^{\beta_R\omega_R(k)}-1}}.
$$

We consider the interaction between the system and the reservoir $R$ of the form
\begin{equation}\label{H_IR}
H_{I,R}=A_{R}|j\rangle\langle i|+A^{*}_{R}|i\rangle\langle j|,\qquad A^{*}_R=\int_{\mathbb{R}^3} g_R(k)a^{*}_R(k)  dk.
\end{equation}

Here $g_R$ is a form factor of interaction and each field $R$ interacts with the corresponding pair of levels
\begin{equation}\label{R}
R={\rm em}, |i\rangle\langle j|=|0\rangle\langle 2|,\qquad R={\rm ph}, |i\rangle\langle j|=|1\rangle\langle 2|,\qquad R={\rm sink}, |i\rangle\langle j|=|0\rangle\langle 1|.
\end{equation}

The total Hamiltonian of the system interacting with three reservoirs is equal to the sum
$$
H=H_S+H_{\rm em}+H_{\rm ph}+H_{\rm sink}+ \lambda\left( H_{I,{\rm em}}+H_{I,{\rm ph}}+H_{I,{\rm sink}}\right)
$$
and acts in the Hilbert space
$$
{\cal H}={\cal H}_S\otimes {\cal H}_{\rm em}\otimes {\cal H}_{\rm ph}\otimes {\cal H}_{\rm sink}.
$$

\medskip

\noindent{\bf Master equation of evolution.}\quad Evolution of the reduced density matrix of the system in the stochastic limit of quantum theory is described by the master equation
\begin{equation}\label{master}
\frac{d}{dt}\rho(t)=\theta\left(\rho(t)\right).
\end{equation}

In the case under consideration the generator $\theta$ of evolution is equal to the sum of three generators for transitions between energy levels
\begin{equation}\label{theta}
\theta=\theta_{\rm sink}+\theta_{\rm ph}+\theta_{\rm em}.
\end{equation}

Generators $\theta_R$ have the form \cite{AcLuVo}
\begin{equation}\label{theta_ij}
\theta_{R}(\rho)=
2\gamma^{-}_{{\rm re},R}
\left(
\langle j|\rho|j\rangle |i\rangle\langle i|
-{1\over 2}
\{\rho,|j\rangle\langle j|\}\right)
-i\gamma^{-}_{{\rm im},R} [\rho,|j\rangle\langle j|]+
$$
$$
+
2\gamma^{+}_{{\rm re},R}
\left(\langle i| \rho |i\rangle |j\rangle\langle j|
-{1\over 2}
\{\rho,|i\rangle\langle i| \}\right)+i\gamma^{+}_{{\rm im},R} [\rho,|i\rangle\langle i|],
\end{equation}
where pairs $|i\rangle$, $|j\rangle$ of levels correspond to reservoirs according to (\ref{R}).

Coefficients in the generator $\theta_R$ take the form
\begin{equation}\label{Regg+1}
\gamma^{+}_{{\rm re}, R}=\pi\int
|g_R(k)|^2\delta(\omega_R(k)-\varepsilon_j+\varepsilon_i)N_R(k)dk,
\end{equation}
\begin{equation}\label{Regg-1}
\gamma^{-}_{{\rm re}, R}=\pi\int
|g_R(k)|^2\delta(\omega_R(k)-\varepsilon_j+\varepsilon_i)(N_R(k)+1)dk,
\end{equation}
\begin{equation}\label{Imgg+1}
\gamma^{+}_{{\rm im}, R}=-\int
|g_R(k)|^2\,{\rm P.P.}\,{1\over\omega_R(k)-\varepsilon_j+\varepsilon_i}N_R(k)dk,
\end{equation}
\begin{equation}\label{Imgg-1}
\gamma^{-}_{{\rm im}, R}=-\int
|g_R(k)|^2\,{\rm P.P.}\,{1\over\omega_R(k)-\varepsilon_j+\varepsilon_i}(N_R(k)+1)dk,
\end{equation}
where the reservoir $R$ is coupled with the transition between energy levels $\varepsilon_i$, $\varepsilon_j$.

For thermal state of the field the above coefficients take form
\begin{equation}\label{gamma-re1}
2\gamma_{{\rm re},R}^{-}={e^{\beta_{R}\varepsilon_j}\over{e^{\beta_{R}\varepsilon_j}-e^{\beta_{R}\varepsilon_i}}}C_{R},
\end{equation}
\begin{equation}\label{gamma+re1}
2\gamma_{{\rm re},R}^{+}={e^{\beta_{R}\varepsilon_i}\over{e^{\beta_{R}\varepsilon_j}-e^{\beta_{R}\varepsilon_i}}}C_{R},
\end{equation}
\begin{equation}\label{C_ij}
C_{R}=2\pi\int |g_R(k)|^2\delta(\omega_R(k)-\varepsilon_j+\varepsilon_i)dk.
\end{equation}
We will call $C_R$ the generalized brightness (for $C_{\rm em}$ this is the brightness of the light, interacting with the system).
In particular,
$$
\frac{\gamma_{{\rm re},R}^{+}}{\gamma_{{\rm re},R}^{-}}=e^{-\beta_R(\varepsilon_j-\varepsilon_i)}.
$$

Moreover $\beta_{\rm sink}$, $\beta_{\rm ph}$, $\beta_{\rm em}$ correspond to temperatures 0Ê, 300Ê and 6000Ê respectively, i.e. the system of excitons interacting with fields is nonequilibrium.  Thus there exists flow of excitons (even in stationary state).  In the present paper we compute this flow.

\section{Nonequilibrium stationary state}\label{Sec3}

In this section we compute the nonequilibrium stationary state of the density matrix and calculate the flow of excitons.

As is known from general theory \cite{AcLuVo}, \cite{notes}, in the stochastic limit the dynamics of the density matrix for non--degenerate system interacting with the environment
can be considered as a combination of dynamics in two subspaces conserved by the evolution: the subspaces of diagonal and off-diagonal matrices. In the diagonal subspace there is convergence to a stationary state (in particular, thermalization for thermal reservoir), and in the off--diagonal subspace there is decoherence (decay of quantum coherences). In the case, when the reservoir is nonequilibrium, the stationary state of the system is also nonequilibrium and there are flows in this subspace, i.e. there is quantum transport.

The diagonal part of the density matrix evolves according to the system of kinetic equations in the Pauli form
$$
\frac{d}{dt}\rho_{ii}(t)=\sum_{j}\left(W_{ij}\rho_{jj}(t)-W_{ji}\rho_{ii}(t)\right),
$$
where $i,j=0,1,2$ and coefficients $W_{ij}$ form a matrix of rates of transitions between the energy levels.

Each summand at the right--hand side of the above system is a flow between the states $i$ and $j$.  Moreover, if the environment is equilibrium (all reservoirs are in thermal states with equal temperatures) then the stationary state of the system will also be thermal with the same temperature. In this case each flow between any pair of states will be zero, this is called the detailed balance case.

For non--equilibrium environment (in particular when we have several reservoirs with different temperatures) the system will converge to nonequilibrium stationary state. In this case  there are non--zero flows in the system. In particular we will be interested in the flow between levels $|1\rangle$ and $|0\rangle$, which corresponds to the exciton absorption rate in the reaction center.

Equations (\ref{master}), (\ref{theta}), (\ref{theta_ij}) imply the following system of equations for the diagonal part of the density matrix
\begin{equation}\label{drho22}
\frac{d\rho_{22}}{dt}=
-2\gamma_{{\rm re},{\rm em} }^{-}\rho_{22}+2\gamma_{{\rm re},{\rm em} }^{+}\rho_{00}
-2\gamma_{{\rm re},{\rm ph} }^{-}\rho_{22}+2\gamma_{{\rm re},{\rm ph} }^{+}\rho_{11};
\end{equation}
\begin{equation}\label{drho11}
\frac{d\rho_{11}}{dt}=
2\gamma_{{\rm re},{\rm ph} }^{-}\rho_{22}-2\gamma_{{\rm re},{\rm ph} }^{+}\rho_{11}
-2\gamma_{{\rm re},{\rm sink} }^{-}\rho_{11}+2\gamma_{{\rm re},{\rm sink} }^{+} \rho_{00};
\end{equation}
\begin{equation}\label{drho00}
\frac{d\rho_{00}}{dt}=
2\gamma_{{\rm re},{\rm em} }^{-}\rho_{22}-2\gamma_{{\rm re},{\rm em} }^{+}\rho_{00}
+2\gamma_{{\rm re},{\rm sink} }^{-}\rho_{11}-2\gamma_{{\rm re},{\rm sink} }^{+}\rho_{00}.
\end{equation}

Let us recall that the sink of excitons corresponds to the reservoir with zero temperature $\beta_{\rm sink} \rightarrow + \infty$, i.e.
\begin{equation}\label{Coeff3_0}
\gamma_{{\rm re},{\rm sink} }^{+}= 0;\qquad
2 \gamma_{{\rm re},{\rm sink} }^{-}=C_{\rm sink} .
\end{equation}
Below we will point out explicitly when we assume $\gamma_{{\rm re},{\rm sink} }^{+}= 0$. The  flow of excitons to the sink becomes $2 \gamma_{{\rm re},{\rm sink} }^{-} \rho_{11}$ in this case.

\begin{Proposition}{\sl 1) Stationary state of the system of equations (\ref{drho22}), (\ref{drho11}), (\ref{drho00}) takes the form
\begin{equation}\label{rho22}
\rho_{22}=\frac{\gamma_{{\rm re},{\rm ph} }^{+} \gamma_{{\rm re},{\rm em} }^{+}+\gamma_{{\rm re},{\rm sink} }^{-}\gamma_{{\rm re},{\rm em} }^{+}+\gamma_{{\rm re},{\rm sink} }^{+}\gamma_{{\rm re},{\rm ph} }^{+}}{\Delta};
\end{equation}
\begin{equation}\label{rho11}
\rho_{11}=\frac{\gamma_{{\rm re},{\rm ph} }^{-}\gamma_{{\rm re},{\rm em} }^{+}+\gamma_{{\rm re},{\rm ph} }^{-}\gamma_{{\rm re},{\rm sink} }^{+}+\gamma_{{\rm re},{\rm em} }^{-}\gamma_{{\rm re},{\rm sink} }^{+}}{\Delta};
\end{equation}
\begin{equation}\label{rho00}
\rho_{00}=\frac{\gamma_{{\rm re},{\rm em} }^{-}\gamma_{{\rm re},{\rm ph} }^{+}+\gamma_{{\rm re},{\rm ph} }^{-}\gamma_{{\rm re},{\rm sink} }^{-}+\gamma_{{\rm re},{\rm em} }^{-}\gamma_{{\rm re},{\rm sink} }^{-}}{\Delta}.
\end{equation}
where
\begin{equation}\label{Delta}
\Delta=
\gamma_{{\rm re},{\rm ph} }^{+}\gamma_{{\rm re},{\rm em} }^{+}+
\gamma_{{\rm re},{\rm sink} }^{-}\gamma_{{\rm re},{\rm em} }^{+}+
\gamma_{{\rm re},{\rm sink} }^{+}\gamma_{{\rm re},{\rm ph} }^{+}+
\gamma_{{\rm re},{\rm ph} }^{-}\gamma_{{\rm re},{\rm em} }^{+}+
$$ $$+
\gamma_{{\rm re},{\rm ph} }^{-}\gamma_{{\rm re},{\rm sink} }^{+}+
\gamma_{{\rm re},{\rm em} }^{-}\gamma_{{\rm re},{\rm sink} }^{+}+
\gamma_{{\rm re},{\rm em} }^{-}\gamma_{{\rm re},{\rm ph} }^{+}+
\gamma_{{\rm re},{\rm ph} }^{-}\gamma_{{\rm re},{\rm sink} }^{-}+
\gamma_{{\rm re},{\rm em} }^{-}\gamma_{{\rm re},{\rm sink} }^{-}.
\end{equation}

2) The flow of excitons to the sink in the above stationary state under assumption $\gamma_{{\rm re},{\rm sink} }^{+}=0$ (at $\beta_{\rm sink}^{-1}\to 0$) equals to
\begin{equation}\label{flow}
{\rm Flow }=\frac{2\gamma_{{\rm re},{\rm sink} }^{-}\gamma_{{\rm re},{\rm ph} }^{-}\gamma_{{\rm re},{\rm em} }^{+}}{\Delta}=
\end{equation}
{\tiny
$$
 =\frac{    C_{\rm ph} C_{\rm em} C_{\rm sink}}
{\left(1+ e^{-\beta_{\rm ph}(\varepsilon_2-\varepsilon_1)} \left(1 + e^{\beta_{\rm em} (\varepsilon_2 - \varepsilon_0)}\right)\right)C_{\rm ph}C_{\rm em}
+\left(e^{\beta_{\rm em}(\varepsilon_2-\varepsilon_0)}-1\right)  C_{\rm ph}C_{\rm sink}+
\left(1+e^{\beta_{\rm em}(\varepsilon_2-\varepsilon_0)}\right)\left(1-e^{-\beta_{\rm ph}(\varepsilon_2-\varepsilon_1)}\right)C_{\rm sink}C_{\rm em}}.
$$
}

}\label{stationary}
\end{Proposition}

\bigskip

\noindent{\bf Remark.} In formula (\ref{flow}) dependence of the flow of excitons to the sink on light brightness $C_{\rm em}$ is line ar for small brightness and is saturated (tends to constant) for high brightnesses. Similar saturation behavior takes place for the dependence of the flow on exciton transport rate $C_{\rm ph}$ and exciton absorption rate in the reaction centre $C_{\rm sink}$.

\section{Relaxation and decoherence}\label{Sec4}

In the previous section we have discussed the form of the stationary state for a system interacting with nonequilibrium environment. In this section we will discuss convergence to the stationary state, i.e. relaxation (in the subspace of diagonal matrices) and decoherence (for the off--diagonal part of the density matrix).

Let us find all the eigenvalues for the matrix of the system (\ref{drho22}), (\ref{drho11}), (\ref{drho00}) of evolution equations for the diagonal part of the density matrix.
Let us consider the characteristic equation
$$
\det
\pmatrix{
-2\gamma_{{\rm re},{\rm em} }^{-}-2\gamma_{{\rm re},{\rm ph} }^{-} - \lambda;& 2\gamma_{{\rm re},{\rm ph} }^{+}; & 2\gamma_{{\rm re},{\rm em} }^{+} \cr
2\gamma_{{\rm re},{\rm ph} }^{-}; & - 2\gamma_{{\rm re},{\rm ph} }^{+}-2\gamma_{{\rm re},{\rm sink} }^{-} - \lambda; & 2\gamma_{{\rm re},{\rm sink} }^{+} \cr
2\gamma_{{\rm re},{\rm em} }^{-}; & 2\gamma_{{\rm re},{\rm sink} }^{-}; & -2\gamma_{{\rm re},{\rm em} }^{+}-2\gamma_{{\rm re},{\rm sink} }^{+} - \lambda}=
$$
$$
= - \lambda \left(\lambda^2+2 \biggl[\gamma_{{\rm re},{\rm em} }^{-}+\gamma_{{\rm re},{\rm ph} }^{-} + \gamma_{{\rm re},{\rm sink} }^{-} + \gamma_{{\rm re},{\rm em} }^{+}+\gamma_{{\rm re},{\rm ph} }^{+} + \gamma_{{\rm re},{\rm sink} }^{+}\biggr]\lambda+
4 \Delta \right) = 0,
$$
where $\Delta$ is defined by (\ref{Delta}).

The roots of this equation are $\lambda=0$ (corresponding to the stationary state) and the roots of the quadratic equation
$$
\lambda^2+2 \biggl[\gamma_{{\rm re},{\rm em} }^{-}+\gamma_{{\rm re},{\rm ph} }^{-} + \gamma_{{\rm re},{\rm sink} }^{-} + \gamma_{{\rm re},{\rm em} }^{+}+\gamma_{{\rm re},{\rm ph} }^{+} + \gamma_{{\rm re},{\rm sink} }^{+}\biggr]\lambda+
4 \Delta =0.
$$

Half-discriminant of the quadratic equation is
\begin{equation}\label{Determinant}
D=\biggl[\gamma_{{\rm re},{\rm em} }^{-}+\gamma_{{\rm re},{\rm ph} }^{-} + \gamma_{{\rm re},{\rm sink} }^{-} + \gamma_{{\rm re},{\rm em} }^{+}+\gamma_{{\rm re},{\rm ph} }^{+} + \gamma_{{\rm re},{\rm sink} }^{+}\biggr]^2-4\Delta.
\end{equation}
Eigenvalues (roots of the quadratic equation)
\begin{equation}\label{lambda_pm}
\lambda_{\pm}=-\biggl[\gamma_{{\rm re},{\rm em} }^{-}+\gamma_{{\rm re},{\rm ph} }^{-} + \gamma_{{\rm re},{\rm sink} }^{-} + \gamma_{{\rm re},{\rm em} }^{+}+\gamma_{{\rm re},{\rm ph} }^{+} + \gamma_{{\rm re},{\rm sink} }^{+}\biggr] \pm \sqrt{D}
\end{equation}
describe the relaxation rate for the diagonal part of the density matrix.

\bigskip

Decoherence is the decay of the off--diagonal part of the density matrix. For three level system under consideration there three (up to Hermitian conjugation) basic matrices in the off--diagonal subspace
\begin{equation}\label{off-diagonal}
|0\rangle\langle 1|,\quad |0\rangle\langle 2|,\quad |1\rangle\langle 2|.
\end{equation}

These matrices are eigenvectors of the evolution generator (\ref{theta})
$$
\theta=\theta_{\rm sink}+\theta_{\rm ph}+\theta_{\rm em},
$$
the details are given in the next proposition.

\begin{Proposition}{\sl
1) The dynamics of diagonal density matrices described by the system of equations (\ref{drho22}), (\ref{drho11}), (\ref{drho00}) reduces to relaxation to the stationary state (\ref{rho22}), (\ref{rho11}), (\ref{rho00}). The relaxation rate is described by the eigenvalues (\ref{lambda_pm})
$$
\lambda_{\pm}=-\biggl[\gamma_{{\rm re},{\rm em} }^{-}+\gamma_{{\rm re},{\rm ph} }^{-} + \gamma_{{\rm re},{\rm sink} }^{-} + \gamma_{{\rm re},{\rm em} }^{+}+\gamma_{{\rm re},{\rm ph} }^{+} + \gamma_{{\rm re},{\rm sink} }^{+}\biggr] \pm \sqrt{D} ,
$$
where $\gamma^{-}_{{\rm re},{\rm ph}}$, $\gamma^{-}_{{\rm re},{\rm em}}$, $\gamma^{+}_{{\rm re},{\rm ph}}$, $\gamma^{-}_{{\rm re},{\rm sink}}$, $\gamma^{+}_{{\rm re},{\rm em}}$, $\gamma^{+}_{{\rm re},{\rm sink}} $ are defined by (\ref{gamma-re1}), (\ref{gamma+re1}), (\ref{C_ij}) and $D$ is given by (\ref{Determinant}).

2) Off diagonal matrices (\ref{off-diagonal}) are eigenvectors for evolution generator (\ref{theta}), i.e.
$$
\theta(|i\rangle\langle j|)= \mu_{ij}|i\rangle\langle j|
$$
and exponentially decay as ${\rm Re}\, \mu_{ij} < 0$. The eigenvalues take the form
\begin{equation}\label{mu_01}
\mu_{01}=
-\gamma^{-}_{{\rm re},{\rm sink}}
-\gamma^{+}_{{\rm re},{\rm sink}}
-\gamma^{+}_{{\rm re},{\rm ph}}
-\gamma^{+}_{{\rm re},{\rm em}}
-i\gamma^{-}_{{\rm im},{\rm sink}}
-i\gamma^{+}_{{\rm im},{\rm sink}}
+i\gamma^{+}_{{\rm im},{\rm ph}}
-i\gamma^{+}_{{\rm im},{\rm em}},
\end{equation}
\begin{equation}\label{mu_02}
\mu_{02}=
-\gamma^{-}_{{\rm re},{\rm em}}
-\gamma^{+}_{{\rm re},{\rm em}}
-\gamma^{+}_{{\rm re},{\rm sink}}
-\gamma^{-}_{{\rm re},{\rm ph}}
-i\gamma^{-}_{{\rm im},{\rm em}}
-i\gamma^{+}_{{\rm im},{\rm em}}
-i\gamma^{+}_{{\rm im},{\rm sink}}
-i\gamma^{-}_{{\rm im},{\rm ph}},
\end{equation}
\begin{equation}\label{mu_12}
\mu_{12}=
-\gamma^{-}_{{\rm re},{\rm ph}}
-\gamma^{+}_{{\rm re},{\rm ph}}
-\gamma^{-}_{{\rm re},{\rm sink}}
-\gamma^{-}_{{\rm re},{\rm em}}
-i\gamma^{-}_{{\rm im},{\rm ph}}
-i\gamma^{+}_{{\rm im},{\rm ph}}
+i\gamma^{-}_{{\rm im},{\rm sink}}
-i\gamma^{-}_{{\rm im},{\rm em}},
\end{equation}
where the coefficients are given by (\ref{Regg+1}), (\ref{Regg-1}), (\ref{Imgg+1}), (\ref{Imgg-1}).
} \label{eigenvalues}
\end{Proposition}

Each eigenvalue contains a contribution $C_{R}$ related to each transition between the energy levels (since $\gamma^{\pm}_{{\rm re},R}$ are proportional to $C_{R}$). Consequently, if at least one of generalized brightnesses $C_{R}$ is high enough then all off--diagonal states decay rapidly. We are interested in coherences related to the transition excited by light, i.e. the matrix element $|0\rangle\langle 2|$.

The rate of convergence to the stationary state is characterized by the minimal value of positive numbers $-{\rm Re}\,(\lambda_{\pm})$. Let us discuss when the numbers $\lambda_{\pm}$ are real and when these numbers are complex and conjugate (since the characteristic equation has real coefficients). To do this we study the sign of the determinant $D$ given by (\ref{Determinant}).

\begin{Proposition}{\sl
Under assumption (\ref{Coeff3_0}) the equation $D \left(C_{\rm ph}, C_{\rm em}, C_{\rm sink}\right) = 0$ with the conditions $C_{\rm em}, C_{\rm sink}, C_{\rm ph}\geq0$ defines a cone with the apex at the point $C_{\rm em}= C_{\rm sink}= C_{\rm ph}=0$.  Inside the cone one has $D \left(C_{\rm ph}, C_{\rm em}, C_{\rm sink}\right)<0$ and outside the cone $D \left(C_{\rm ph}, C_{\rm em}, C_{\rm sink}\right)>0$. The cone is tangent to the plane $C_{\rm ph} = 0$ by the half-line
\begin{equation}
C_{\rm ph} =0, \quad C_{\rm sink} = \frac{e^{\beta_{\rm em}(\varepsilon_2-\varepsilon_0)}+1}{e^{\beta_{\rm em}(\varepsilon_2-\varepsilon_0)}-1}C_{\rm em},
\end{equation}
and is tangent to the plane $\gamma^{+}_{{\rm re},{\rm ph}}=\gamma^{+}_{{\rm re},{\rm em}}$ by the half-line
\begin{equation}
C_{\rm ph} = \frac{e^{\beta_{\rm ph}(\varepsilon_2-\varepsilon_1)}-1}{e^{\beta_{\rm em}(\varepsilon_2-\varepsilon_0)}-1} C_{\rm em} , \quad C_{\rm sink} = \frac{e^{\beta_{\rm ph}(\varepsilon_2-\varepsilon_1)} + e^{\beta_{\rm em}(\varepsilon_2-\varepsilon_0)}}{e^{\beta_{\rm em}(\varepsilon_2-\varepsilon_0)}-1}C_{\rm em}.
\end{equation}
}\label{cone}
\end{Proposition}

\noindent{\textbf{Remark.}} Here and below we use the word ''cone'' to denote a part of a cone surface lying at the same side of the apex.

\medskip

\noindent{\textbf{Corollary.}} The convergence rate (in the limit $t \rightarrow + \infty$) to the stationary state is defined by $|{\rm Re}\, \lambda_-|$ and equals
\begin{equation}
|{\rm Re}\, \lambda_-| = \left\{
\begin{array}{c}
\gamma^{-}_{{\rm re},{\rm ph}}+\gamma^{-}_{{\rm re},{\rm em}}+\gamma^{+}_{{\rm re},{\rm ph}}+\gamma^{-}_{{\rm re},{\rm sink}}+\gamma^{+}_{{\rm re},{\rm em}}, \qquad \texttt{inside the cone}\\
\gamma^{-}_{{\rm re},{\rm ph}}+\gamma^{-}_{{\rm re},{\rm em}}+\gamma^{+}_{{\rm re},{\rm ph}}+\gamma^{-}_{{\rm re},{\rm sink}}+\gamma^{+}_{{\rm re},{\rm em}}- \sqrt{D}, \qquad \texttt{outside the cone}
\end{array} \right.
\end{equation}
The function  $|{\rm Re}\, \lambda_-|$ is continuous with respect to $C_{R}$.

\medskip

\noindent{\textbf{Remark.}} The cone $D \left(C_{\rm ph}, C_{\rm em}, C_{\rm sink}\right) = 0$ is tangent to one of three planes $C_{R}=0$ and intersects none of them. Thus in the case of only two reservoirs the discriminant $D$ is always non--negative. Therefore at least three reservoirs are necessary for complex roots to appear.

\medskip

\noindent{\bf Comparison of the relaxation rate for diagonal matrix elements with decoherence rate.}
For non--degenerate system under consideration the slowest rate of decoherence is obtained if $C_{\rm em}=0$ (i.e. the light is switched off).
The exciton transport and absorption cannot be switched off.
Let us find the bounds for the relaxation rate to the decoherence rate ratio.

\begin{Proposition}{\sl
Under assumption (\ref{Coeff3_0}), inside the cone $D \left(C_{\rm ph}, C_{\rm em}, C_{\rm sink}\right) = 0, \gamma^{-}_{{\rm re},{\rm ph}}\geq0$  the ratio $\frac{|{\rm Re} \, \lambda_{\pm}|}{|{\rm Re} \,\mu_{02}| } $ is bounded by
\begin{equation}
\frac{|{\rm Re}\, \lambda_{\pm}|}{|{\rm Re} \,\mu_{02}| }  <2+ \sqrt{\frac{e^{\beta_{\rm ph} (\varepsilon_2 - \varepsilon_1)}}{\left(1+ e^{\beta_{\rm em} (\varepsilon_2 - \varepsilon_0)} \right)\left(1+e^{\beta_{\rm ph} (\varepsilon_2 - \varepsilon_1)}+ e^{\beta_{\rm em} (\varepsilon_2 - \varepsilon_0)} \right)}},
\end{equation}
\begin{equation}
\frac{|{\rm Re} \,\lambda_{\pm}|}{|{\rm Re} \,\mu_{02}| }  > 2- \sqrt{\frac{e^{\beta_{\rm ph} (\varepsilon_2 - \varepsilon_1)}}{\left(1+ e^{\beta_{\rm em} (\varepsilon_2 - \varepsilon_0)} \right)\left(1+e^{\beta_{\rm ph} (\varepsilon_2 - \varepsilon_1)}+ e^{\beta_{\rm em} (\varepsilon_2 - \varepsilon_0)} \right)}}.
\end{equation}

The bounds are reached on the cone surface $D \left(C_{\rm ph}, C_{\rm em}, C_{\rm sink}\right) = 0$ on the half-lines
$$
\frac{C_{\rm em}}{C_{\rm ph}} = \frac{e^{\beta_{\rm em}(\varepsilon_2 - \varepsilon_0)} -1}{e^{\beta_{\rm ph}(\varepsilon_2 - \varepsilon_1)} -1}
\left(2 + \frac{e^{\beta_{\rm ph} (\varepsilon_2 - \varepsilon_1)}}{1+ e^{\beta_{\rm em} (\varepsilon_2 - \varepsilon_0)}}\right),
$$
\small
$$
\frac{C_{\rm sink}}{C_{\rm ph}} = \frac{1}{e^{\beta_{\rm ph}(\varepsilon_2 - \varepsilon_1)} -1}
\left(1+ 2\left(e^{\beta_{\rm ph}(\varepsilon_2 - \varepsilon_1)} + e^{\beta_{\rm em}(\varepsilon_2 - \varepsilon_0)} \right)
\pm 2\sqrt{\frac{e^{\beta_{\rm ph} (\varepsilon_2 - \varepsilon_1)}\left(1+e^{\beta_{\rm ph} (\varepsilon_2 - \varepsilon_1)}+ e^{\beta_{\rm em} (\varepsilon_2 - \varepsilon_0)} \right)}{1+ e^{\beta_{\rm em} (\varepsilon_2 - \varepsilon_0)}}}\right).
$$
\normalsize
}\label{relaxtodecoh}
\end{Proposition}

\noindent{\textbf{Remark.}} Taking the typical values
\begin{equation}\label{typical}
\beta_{\rm em}^{-1}=6000K, \quad \beta_{\rm ph}^{-1}=300K, \quad \varepsilon_2 - \varepsilon_0 = 2 eV, \quad \varepsilon_2 - \varepsilon_1 = 0.5 eV
\end{equation}
we get the numerical estimates
$$
\frac{C_{\rm em}}{C_{\rm ph}}\approx 0.96 \quad \frac{C_{\rm sink}}{C_{\rm ph}} \approx 2.26, \quad \frac{|{\rm Re}\, \lambda_{\pm}|}{|{\rm Re}\, \mu_{02}| } \approx 2.13.
$$

\begin{Proposition}{\sl
The global maximum value of the ratio $\frac{|{\rm Re}\, \lambda_{+}|}{|{\rm Re}\, \mu_{02}| }$ is given by
\begin{equation}\label{globaleq}
\frac{|{\rm Re}\, \lambda_{+}|}{|{\rm Re}\, \mu_{02}| }  =2+ \sqrt{\frac{e^{\beta_{\rm ph} (\varepsilon_2 - \varepsilon_1)}}{\left(1+ e^{\beta_{\rm em} (\varepsilon_2 - \varepsilon_0)} \right)\left(1+e^{\beta_{\rm ph} (\varepsilon_2 - \varepsilon_1)}+ e^{\beta_{\rm em} (\varepsilon_2 - \varepsilon_0)} \right)}}.
\end{equation}
}\label{globalmax}
\end{Proposition}

\noindent{\textbf{Remark.}} One has the estimate for (\ref{globaleq}) under assumption (\ref{Coeff3_0})
$$
\frac{|{\rm Re}\, \lambda_{+}|}{|{\rm Re}\, \mu_{02}| }  < 2 + \frac{\sqrt{2}}{2} \approx 2.71
$$
for all brightnesses and temperatures.

As the upper bound of the relaxation rate to the decoherence rate ratio appeared to be more than 1 and the ratio is a monotonic function with respect to  $\gamma^{-}_{{\rm re},{\rm sink}}$ outside the cone, then it is natural to ask the question, under what conditions  the relaxation rate and decoherence rate are equal? The answer is given by the following proposition.

\begin{Proposition}{\sl
The relaxation and decoherence rates are equal on a part of the cone surface, which can be given by single-valued function $C_{\rm sink}$ depending on $C_{\rm ph}$ and $C_{\rm em}$ in the brightnesses range excluding the line $C_{\rm sink}= C_{\rm em}$
\tiny
\begin{equation}\label{cone2}
\gamma^{-}_{{\rm re},{\rm sink}} =\frac{(\gamma^{-}_{{\rm re},{\rm ph}})^2+2 \gamma^{-}_{{\rm re},{\rm ph}} \gamma^{-}_{{\rm re},{\rm em}}+2 \gamma^{-}_{{\rm re},{\rm ph}} \gamma^{+}_{{\rm re},{\rm ph}} -2 \gamma^{-}_{{\rm re},{\rm ph}} \gamma^{+}_{{\rm re},{\rm em}}+(\gamma^{-}_{{\rm re},{\rm em}})^2-2 \gamma^{-}_{{\rm re},{\rm em}} \gamma^{+}_{{\rm re},{\rm ph}}+2 \gamma^{-}_{{\rm re},{\rm em}} \gamma^{+}_{{\rm re},{\rm em}}-2 \gamma^{+}_{{\rm re},{\rm ph}} \gamma^{+}_{{\rm re},{\rm em}}+(\gamma^{+}_{{\rm re},{\rm em}})^2}{2 (\gamma^{-}_{{\rm re},{\rm ph}}+ \gamma^{-}_{{\rm re},{\rm em}}+ \gamma^{+}_{{\rm re},{\rm em}})}
\end{equation}
\normalsize
The line $C_{\rm sink}= C_{\rm em}$ belongs to the cone. The cone $D \left(C_{\rm ph}, C_{\rm em}, C_{\rm sink}\right) = 0$ lies inside the cone (\ref{cone2}).
}\label{eqreldech}
\end{Proposition}

Since the case $D<0$ is possible, one can ask is it possible to obtain the oscillatory behavior? The answer is no.
Let us define the oscillation quality inside the cone
\begin{equation}
Q = \frac{\sqrt{-D}}{\gamma^{-}_{{\rm re},{\rm ph}}+\gamma^{-}_{{\rm re},{\rm em}}+\gamma^{+}_{{\rm re},{\rm ph}}+\gamma^{-}_{{\rm re},{\rm sink}}+\gamma^{+}_{{\rm re},{\rm em}}}
\end{equation}
At large values $Q/\pi$ describes the number of oscillations during the relaxation time, so it is natural to interpret
$$
Q > \pi
$$
as a condition for transition to oscillatory behavior.

\begin{Proposition}{\sl
Under assumption (\ref{Coeff3_0}), the maximum oscillations quality inside the cone equals
\begin{equation}\label{quality}
\left.Q\right|_{\max} = \sqrt{\frac{e^{\beta_{\rm ph}(\varepsilon_2 - \varepsilon_1)} }{e^{\beta_{\rm ph}(\varepsilon_2 - \varepsilon_1)}  (4 e^{\beta_{\rm em}(\varepsilon_2 - \varepsilon_0)} +3)+4 (e^{\beta_{\rm em}(\varepsilon_2 - \varepsilon_0)} +1)^2}}
\end{equation}
and is reached on a half-line
\begin{equation}
\frac{\gamma_{{\rm re}, {\rm em}}^+}{\gamma_{{\rm re}, {\rm ph}}^+} = 2 + \frac{e^{\beta_{\rm ph}(\varepsilon_2 - \varepsilon_1)}}{1 + e^{\beta_{\rm em}(\varepsilon_2 - \varepsilon_0)}}
\end{equation}
\begin{equation}
\frac{\gamma_{{\rm re}, {\rm sink}}^-}{\gamma_{{\rm re}, {\rm ph}}^+} = \left(1 + \frac{e^{\beta_{\rm ph}(\varepsilon_2 - \varepsilon_1)}}{1 + e^{\beta_{\rm em}(\varepsilon_2 - \varepsilon_0)}} \right)\left(1 + \frac{e^{\beta_{\rm em}(\varepsilon_2 - \varepsilon_0)}}{1 + e^{\beta_{\rm em}(\varepsilon_2 - \varepsilon_0)}} \right)
\end{equation}
} \label{maxquality}
\end{Proposition}

\noindent{\textbf{Remark.}}  If one substitutes (\ref{typical}) then
$$
\left.Q\right|_{\max} \approx 0.067.
$$
If one maximizes (\ref{quality}) with respect to temperatures, then
$$
\left.Q\right|_{\max} = \frac{1}{\sqrt{7}} \approx 0.38.
$$
Thus $Q < \pi$ and one can interpret the result as absence of oscillatory dependence on time.

\medskip

\noindent{\bf Comparison of the flow of excitons to the sink with the decoherence.}\quad To test if the coherence is preserved during the time exciton needs to be absorbed in the reaction centre one should check if the ratio of the flow to decoherence rates is greater than 1.

\begin{Proposition}{\sl The exciton flow to sink to decoherence rate ratio has the upper bound
\begin{equation}\label{flowtodecoheq}
\frac{\rm Flow}{|{\rm Re}\, \mu_{02}|} < \frac{1}{2 \left(1 + e^{\beta_{\rm em}(\varepsilon_2 - \varepsilon_0)}\right)}.
\end{equation}
}\label{flowtodecoh}
\end{Proposition}

\noindent{\textbf{Remark.}} At (\ref{typical}) the bound equals
$$
\left.\frac{\rm Flow}{|{\rm Re}\, \mu_{02}|}\right|_{\max}  \approx 0.009.
$$
If one maximizes (\ref{flowtodecoheq}) with respect to temperature then
$$
\frac{\rm Flow}{|{\rm Re}\, \mu_{02}|} < \frac{1}{4} = 0.25.
$$

Thus coherence is not preserved at the transport and absorption time scale.

\section{Interaction with coherent field}\label{Sec5}

Interaction of atom with coherent field (with application of coherent states of the field) in the stochastic limit approach was considered in \cite{lambdaatom}. In this section we will consider a simpler, but equivalent approach. We will add a classical term in special scaling limit to the quantum field of the reservoir.

Let us recall that the Hamiltonian of interaction of the system with light has the form (\ref{H_IR}) (for $R={\rm em}$).
We introduce the external coherent field adding a classical contribution of the form
$$
A_{\rm em}^{*}\mapsto A_{\rm em}^{*} + \lambda s(\lambda^2 t)e^{it\omega_{\rm em}},
$$
i.e. an oscillating classical field multiplied by the term $s(\lambda^2 t)$ (signal envelope, a complex valued function)  varying slowly with respect to time is added to the quantum field. Here $\omega_{\rm em}=\varepsilon_2-\varepsilon_0$ is a Bohr frequency of the transition between states $|0\rangle$ and $|2\rangle$ (the classical field is in resonance with the transition).

This contribution can be non--zero in the stochastic limit since the oscillating exponents arising from the free evolution of the interaction Hamiltonian in the stochastic limit are canceled. The free evolution of the classical field contribution is constant and after the time rescaling $t\mapsto t/\lambda^2$ is equal to
$$
H_{\rm eff}(t)= \overline{s(t)}|2\rangle\langle 0|+s(t)|0\rangle\langle 2|.
$$

This gives the following proposition.

\begin{Proposition}{\sl
The evolution equation of the reduced density matrix of the system (described in section \ref{the_system}) in the stochastic limit in presence of an electromagnetic coherent field takes the form
\begin{equation}\label{generator}
\frac{d}{dt}\rho(t)=i[\rho(t), H_{\rm eff}(t)]+\theta(\rho(t)),
\end{equation}
where $\theta$ is the Lindblad generator (\ref{theta}) for quantum fields in a Gaussian state, the effective Hamiltonian for the considered system takes the form
$$
H_{\rm eff}(t)= \overline{s(t)}|2\rangle\langle 0|+s(t)|0\rangle\langle 2|.
$$
}\label{genwithfield}
\end{Proposition}

\noindent{\bf The generator of evolution of the reduced density matrix.}\quad
Let us consider the matrix representation of the generator (\ref{generator}) (where $s={\rm const}$) in the basis
\begin{equation}\label{basis}
|2\rangle\langle 2|, \quad |1\rangle\langle 1|, \quad |0\rangle\langle 0| , \quad |2\rangle\langle 0|, \quad |0\rangle\langle 2|, \quad |0\rangle\langle 1| , \quad |2\rangle\langle 1|, \quad |1\rangle\langle 0|  , \quad  |1\rangle\langle 2|.
\end{equation}
This representation is given by the following proposition
\begin{Proposition}{\sl
The generator
\begin{equation}\label{generator_L}
L=\theta+i[\cdot, H_{\rm eff}]
\end{equation}
in the basis (\ref{basis}) has a matrix of the block--diagonal form
$$
L =
\pmatrix{
L_{ds} & 0 & 0\cr
0 & L_{nd} & 0\cr
0 & 0 & L_{nd}^*
}
$$
where
\begin{equation}\label{L_ds}
L_{ds}=\pmatrix{
-2\gamma_{{\rm re},{\rm em} }^{-}-2\gamma_{{\rm re},{\rm ph} }^{-};& 2\gamma_{{\rm re},{\rm ph} }^{+}; & 2\gamma_{{\rm re},{\rm em} }^{+};   &       is;&      -is\cr
2\gamma_{{\rm re},{\rm ph} }^{-}; & - 2\gamma_{{\rm re},{\rm ph} }^{+}-2\gamma_{{\rm re},{\rm sink} }^{-}; & 2\gamma_{{\rm re},{\rm sink} }^{+};&        0;&        0\cr
2\gamma_{{\rm re},{\rm em} }^{-}; & 2\gamma_{{\rm re},{\rm sink} }^{-}; & -2\gamma_{{\rm re},{\rm em} }^{+}-2\gamma_{{\rm re},{\rm sink} }^{+};&      -is;&       is\cr
  is;&    0;&  -is;& \mu_{20};&        0\cr
 -is;&    0;&   is;&        0;& \mu_{02} \cr
}
\end{equation}
\begin{equation}\label{L_nd}
L_{nd} =\pmatrix{
 \mu_{01};&      -is\cr
 -is;& \mu_{21}&\cr
}\quad
\end{equation}
}\label{generatorwithfield}
\end{Proposition}

Here $\gamma^{\pm}_{\rm re}$ are given by (\ref{gamma-re1}), (\ref{gamma+re1}) and $\mu_{ij}$ are eigenvalues of the operator $\theta$ for the off-diagonal vectors $|i\rangle\langle j|$ given by (\ref{mu_01}), (\ref{mu_02}), (\ref{mu_12})
$$
\theta(|i\rangle\langle j|)=\mu_{ij}|i\rangle\langle j|.
$$

\medskip

\noindent{\bf Stationary states.}\quad It is easy to see that the matrix elements $\rho_{10}$, $\rho_{01}$, $\rho_{21}$, $\rho_{12}$ decay exponentially (because the corresponding eigenvalues $\mu_{ij}$ have negative real parts). Thus the stationary state for the generator $L$ belongs to the subspace corresponding to the first five lines of the above matrix (which contains all diagonal matrix elements and two off--diagonal matrix elements $\rho_{20}$ and $\rho_{02}$). In this subspace the equation for the stationary state takes the form
$$
\pmatrix{
-2\gamma_{{\rm re},{\rm em} }^{-}-2\gamma_{{\rm re},{\rm ph} }^{-};& 2\gamma_{{\rm re},{\rm ph} }^{+}; & 2\gamma_{{\rm re},{\rm em} }^{+};   &       is;&      -is\cr
2\gamma_{{\rm re},{\rm ph} }^{-}; & - 2\gamma_{{\rm re},{\rm ph} }^{+}-2\gamma_{{\rm re},{\rm sink} }^{-}; & 2\gamma_{{\rm re},{\rm sink} }^{+};   &        0;&        0\cr
2\gamma_{{\rm re},{\rm em} }^{-}; & 2\gamma_{{\rm re},{\rm sink} }^{-}; & -2\gamma_{{\rm re},{\rm em} }^{+}-2\gamma_{{\rm re},{\rm sink} }^{+};&      -is;&       is\cr
  is;&    0;&  -is;& \mu_{20};&        0\cr
 -is;&    0;&   is;&        0;& \mu_{02}\cr}
\pmatrix{\rho_{22}\cr \rho_{11}\cr \rho_{00}\cr \rho_{20}\cr \rho_{02}\cr }=0.
$$

It is easy to see that in the stationary state
$$
\rho_{20}=-\frac{is}{\mu_{20}}(\rho_{22}-\rho_{00});\qquad
\rho_{02}=\frac{is}{\mu_{02}}(\rho_{22}-\rho_{00}).
$$

Substituting these expressions into the equations for the diagonal part of the density matrix, we obtain the system of linear equations for the stationary state
$$
\pmatrix{
-2\gamma_{{\rm re},{\rm em} }^{-}-2\gamma_{{\rm re},{\rm ph} }^{-}+\frac{s^2}{\mu_{20}}+\frac{s^2}{\mu_{02}};&    2\gamma_{{\rm re},{\rm ph} }^{+};&  2\gamma_{{\rm re},{\rm em} }^{+}-\frac{s^2}{\mu_{20}}-\frac{s^2}{\mu_{02}}\cr
2\gamma_{{\rm re},{\rm ph} }^{-}; & - 2\gamma_{{\rm re},{\rm ph} }^{+}-2\gamma_{{\rm re},{\rm sink} }^{-}; & 2\gamma_{{\rm re},{\rm sink} }^{+}\cr
2\gamma_{{\rm re},{\rm em} }^{-}-\frac{s^2}{\mu_{20}}-\frac{s^2}{\mu_{02}};& 2\gamma_{{\rm re},{\rm sink} }^{-};& -2\gamma_{{\rm re},{\rm em} }^{+}-2\gamma_{{\rm re},{\rm sink} }^{+}+\frac{s^2}{\mu_{20}}+\frac{s^2}{\mu_{02}}\cr}
\pmatrix{\rho_{22}\cr \rho_{11}\cr \rho_{00}\cr }=0.
$$

Eigenvalues of this matrix are 0, $\lambda_{\pm}(s)$, where $\lambda_{\pm}(s)$ are deformations of $\lambda_{\pm}$ given by  (\ref{lambda_pm}).

Comparison of this system of equations and the system (\ref{drho22}), (\ref{drho11}, (\ref{drho00}) implies the following proposition.

\begin{Proposition}
{\sl The stationary state for the generator (\ref{generator_L}) in the presence of external coherent field (where $s={\rm const}$ is real) exists, is unique and has the following form:

The diagonal elements of the density matrix are described by expressions (\ref{rho22}), (\ref{rho11}), (\ref{rho00}) where the following substitution was made
\begin{equation}\label{change1}
\gamma_{{\rm re},{\rm em} }^{-} \mapsto \gamma_{{\rm re},{\rm em} }^{-}-s^2{\rm Re}\frac{1}{\mu_{20}},
\end{equation}
\begin{equation}\label{change2}
\gamma_{{\rm re},{\rm em} }^{+}\mapsto \gamma_{{\rm re},{\rm em} }^{+} -s^2{\rm Re}\frac{1}{\mu_{20}}.
\end{equation}

The off--diagonal elements in the stationary state are
$$
\rho_{20}=\frac{is}{\mu_{20}}\frac{(\gamma_{{\rm re},{\rm ph} }^{+}+\gamma_{{\rm re},{\rm sink} }^{-})(\gamma_{{\rm re},{\rm em} }^{-}-\gamma_{{\rm re},{\rm em} }^{+})-\gamma_{{\rm re},{\rm sink} }^{+}\gamma_{{\rm re},{\rm ph} }^{+}+\gamma_{{\rm re},{\rm ph} }^{-}\gamma_{{\rm re},{\rm sink} }^{-}}{\Delta'},
$$
$$
\rho_{02}=\rho_{20}^{*},
$$
where $\Delta'$ is obtained from (\ref{Delta}) by the substitution (\ref{change1}), (\ref{change2}).
}\label{statstatefield}
\end{Proposition}

The quantity $2(\gamma_{{\rm re},{\rm em} }^{-}-\gamma_{{\rm re},{\rm em} }^{+})$ is the photon spontaneous emission rate.

\medskip

\noindent{\bf Convergence to the stationary state.}\quad To study convergence to the stationary state let us discuss eigenvalues of the matrix $L_{ds}$ given by (\ref{L_ds}). The characteristic equation
\begin{equation}\label{char_eq_det}
\det (\lambda I - L_{ds})=0
\end{equation}
takes the form
\begin{equation}\label{char_eq_s}
\lambda(h_1(\lambda) + s^2 h_2(\lambda))=0,
\end{equation}
where
\begin{equation}\label{h1_def}
h_1(\lambda) = (\lambda- \lambda_+)(\lambda- \lambda_-)(\lambda- \mu_{02})(\lambda- \overline{\mu_{02}})
\end{equation}

\begin{equation}\label{h2_def}
h_2(\lambda) = 2 (2\lambda - \mu_{02} - \overline{\mu_{02}})(\lambda - \lambda_r), \quad \lambda_r = -\biggl[\gamma_{{\rm re},{\rm ph}}^-  +2 \gamma_{{\rm re},{\rm ph}}^+  +\gamma_{{\rm re},{\rm sink}}^+  +2 \gamma_{{\rm re},{\rm sink}}^-\biggr]
\end{equation}
are independent of the parameter $s$, $\lambda_{\pm}$  are defined by (\ref{lambda_pm}), $\mu_{02}$ is defined by (\ref{mu_02}). In particular the characteristic equation without the coherent field is $\lambda h_1(\lambda) =0$.

Let us discuss the possibility of pure oscillatory regime for the above system of equations.

\begin{Proposition}{\sl The necessary condition for the presence of pure imaginary (non-zero) roots of the 4-th order equation
\begin{equation}\label{quartic_eq}
\lambda^4 + w_3 \lambda^3 + w_2 \lambda^2 + w_1 \lambda + w_0 = 0
\end{equation}
with real coefficients $w_0, w_1, w_2, w_3$ has the form
\begin{equation}\label{im_cond}
w_1^2 - w_1 w_2 w_3 +  w_3^2 w_0 = 0
\end{equation}
}\label{imrootcond}
\end{Proposition}

This implies the following.

\begin{Proposition}{\sl The equation (\ref{char_eq_s}) can have pure imaginary roots only if $\gamma_{{\rm re},{\rm em} }^{-}=\gamma_{{\rm re},{\rm ph} }^{-} = 0$.
}\label{noosc}
\end{Proposition}

This regime is possible only when emission and absorption of photons (including spontaneous emission) and interaction with phonons are switched off. In the non--trivial regimes the corresponding eigenvalues have negative real part. Therefore at large times the reduced density matrix of the system converges to the stationary state described by proposition \ref{statstatefield}.

\medskip

\noindent{\bf Behavior in weak and strong fields.}\quad
Let us compute the asymptotics for small $s$ of the eigenvalues of the evolution generator. This asymptotics may be computed using the exact formulas for the eigenvalues as solutions of (\ref{char_eq_s}).

\begin{Proposition}{\sl (weak field) The asymptotics of the eigenvalues of the matrix $L_{ds}$ in the regime of small $s$ is given by the expressions

\begin{equation}\label{pertpm}
\lambda_{\pm}(s)= \lambda_{\pm}\mp 4 \frac{(\lambda_{\pm} - \lambda_r)(\lambda_{\pm} - {\rm Re}\,\mu_{02} )}{(\lambda_+ - \lambda_-)(\lambda_{\pm} - \mu_{02})(\lambda_{\pm} - \overline{\mu}_{02})} s^2 + O(s^4)
\end{equation}

\begin{equation}\label{pert02}
\mu_{02}(s) =\mu_{02}- 2 \frac{\mu_{02} - \lambda_r}{(\mu_{02} - \lambda_+)(\mu_{02} - \lambda_-)} s^2 + O(s^4), \qquad  \mu_{20}(s)= \overline{\mu}_{02}(s)
\end{equation}
\label{weakfield}
}
\end{Proposition}

Here we denoted the eigenvalues of the matrix $L_{ds}$  by $\lambda_{\pm}(s)$, $\lambda_{\pm}(0) = \lambda_{\pm}$;  $\mu_{02}(s)$ and $\mu_{20}(s)$, $\mu_{02}(0) = \mu_{02}$, $\mu_{20}(0) = \mu_{20}$; these eigenvalues are deformations of the corresponding eigenvalues for the case of non--zero coherent field.

\medskip

\noindent{\textbf{Remark.}}
For non--zero $s$ the eigenvalues $\lambda_{\pm}$ remain real outside the cone $D=0$ given by (\ref{Determinant}).

\medskip

Let us discuss the dynamics of the density matrix of the system in strong coherent field (for large $s$).
The matrix $L_{ds}$ possesses five eigenvalues, one of the eigenvalues is zero, the other we denote $\lambda_s(s), \overline{\lambda_s}(s), \lambda_r(s), \lambda_{\mu}(s)$ (two of the eigenvalues are mutually conjugated). Limits of these eigenvalues for $s\to \infty$ we denote $\lambda_s, \overline{\lambda_s}, \lambda_r, \lambda_{\mu}$.

\begin{Proposition}{\sl (strong field) The eigenvalues $\lambda_s(s), \overline{\lambda_s}(s), \lambda_r(s), \lambda_{\mu}(s)$ of the matrix  $L_{ds}$ possess the following asymptotic for $s \rightarrow \infty$:
\begin{equation}
\lambda_r(s) = \lambda_r + O\left(\frac{1}{s}\right),
\end{equation}
\begin{equation}
\lambda_{\mu}(s) = {\rm Re}\, \mu_{02} + O\left(\frac{1}{s}\right),
\end{equation}
\begin{equation}
\lambda_s(s) = 2 i s - \frac12 \left(\gamma_{{\rm re}, {\rm ph}}^- + 2\gamma_{{\rm re}, {\rm em}}^-  + 2 \gamma_{{\rm re}, {\rm em}}^+ + \gamma_{{\rm re}, {\rm sink}}^+\right) + O\left(\frac{1}{s}\right).
\end{equation}
}\label{strongfield}
\end{Proposition}

\noindent{\bf Decoherence in external coherent field.}\quad For non--zero laser field (when $s\ne 0$) the coherences given by matrix elements $\rho_{02}$, $\rho_{20}$ are non--zero for the stationary state. Let us discuss the dependence on $s$ of the decay for matrix elements $\rho_{21}$, $\rho_{01}$.

\begin{Proposition}{\sl Eigenvalues of the matrix $L_{nd}$ given by (\ref{L_nd}) possess the following dependence on $s$
\begin{equation}\label{mu21}
\mu_{21} (s)= \frac12 \left(\mu_{21} + \mu_{01} + \sqrt{(\mu_{21} - \mu_{01} )^2 -4 s^2} \right)
\end{equation}
\begin{equation}\label{mu01}
\mu_{01} (s)= \frac12 \left(\mu_{21} + \mu_{01} - \sqrt{(\mu_{21} - \mu_{01} )^2 -4 s^2} \right)
\end{equation}
Here $\mu_{01}$, $\mu_{21}$ are given by (\ref{mu_01}), (\ref{mu_12}).
}\label{dechextfield}
\end{Proposition}

\noindent{\textbf{Remark.}} The sum of the decoherence rates $-{\rm Re}\,\mu_{21} (s)-{\rm Re}\, \mu_{01} (s) = - {\rm Re}\,(\mu_{21} + \mu_{01})$ does not depend on the external field. Thus the external coherent field ''redistributes'' decoherence rates for the transitions between the pairs of levels $|1\rangle$, $|0\rangle$ and $|1\rangle$, $|2\rangle$. In the limit $s \rightarrow \infty$ the decoherence rates become equal: $-{\rm Re}\,\mu_{21} (\infty) =-{\rm Re}\, \mu_{01} (\infty)$.

\section{Conclusion}

In the present paper we have discussed interaction of a three level system with nonequilibrium environment (given by three different reservoirs) and applications to quantum photosynthesis. The results are as follows.

Proposition \ref{stationary} of Section \ref{Sec3} describes the nonequilibrium stationary state for the system with nonequilibrium environment and describes the dependence of the current (the flow of excitons for the case of a model of quantum photosynthesis) on the environment. We found that our model predicts a linear behavior for low brightness of the light which saturates (tends to constant) for hight brightness (and similar behavior with respect to properties of other reservoirs, called the generalized brightnesses, proportional to the number of quanta of the reservoir fields).

Section \ref{Sec4} discusses relaxation and decoherence to the stationary state. It is shown that the relaxation includes not only damping but also oscillations and the relaxation and decoherence are sufficiently fast.

In Section \ref{Sec5} these results are generalized for the case when the light field contains a coherent component which is in resonance with the transition. We show that interaction with a coherent field leads to excitation of quantum coherences (off--diagonal elements of the density matrix) and discuss the form of the corresponding stationary state, decoherence and relaxation.

It would be interesting to study also a relation wih quantum control theory \cite{Pec}.

\bigskip

\noindent{\bf Acknowledgements.}\quad This work is supported by the Russian Science Foundation under grant 14-11-00687 and
performed in the Steklov Mathematical Institute of Russian Academy of Sciences, Moscow, Russia.

\end{document}